\documentclass{article}
\usepackage{fullpage}
\usepackage{authblk}
\usepackage{amsmath}
\usepackage{amssymb}
\usepackage{graphicx}
\usepackage[colorlinks=true]{hyperref}
\usepackage[font=small,labelfont=bf]{caption}
\usepackage{booktabs}

\begin{document}

\title {Scalable CFD Simulations in Multi-Billion Voxel Micro-CT Images of Porous Materials Using OpenFOAM on ARCHER2}                      

\author[1]{Julien Maes}
\author[2]{Gavin J. Pringle}
\author[1]{Hannah P. Menke}

\affil[1]{Institute of GeoEnergy Engineering, Heriot-Watt University, Edinburgh, U.K.}
\affil[2]{EPCC, University of Edinburgh,United Kingdom}

\maketitle

\begin{abstract}
This study investigates the use of High-Performance Computing (HPC) to simulate flow and transport in ultra-large micro-CT images of porous materials using Computational Fluid Dynamics (CFD). Two distinct rock samples, representative of two different rock formations - Bentheimer sandstone and Estaillades carbonate - are investigated. The Bentheimer sandstone image, with dimensions 1,950×1,950×10,800 voxels at 6 $\mu$m resolution, comprising 41 billion voxels, represents a largely homogeneous structure, while the Estaillades carbonate image, at 1,144×1,144×6,000 voxels and 3.9676 $\mu$m resolution, amounting to 8 billion voxels, features greater heterogeneity, including micro-porous regions. These images are used for direct CFD simulations with GeoChemFoam, our OpenFOAM-based numerical solver, leveraging the computational resources of the UK supercomputer ARCHER2. One of the key aspects of the study is the use of the Darcy-Brinkman-Stokes approach, for which the solid surface is represented using a volumetric indicator function, rather than a complex mesh. This enables the use of simple Cartesian meshes that can be generated in parallel in an efficient and scalable manner. The study explores both weak and strong scaling through subvolume decomposition, demonstrating that, due to the strong scalability and the computational power of ARCHER2, full-resolution CFD simulations can be carried out without the need for image size reduction. This work illustrates the potential of HPC to perform detailed, full-scale simulations on large, high-resolution micro-CT data. The approach relies on a meshing strategy that leverages simple, parallelisable Cartesian grids derived from volumetric indicator functions, eliminating the need for complex surface-conforming meshes and allowing scalable simulation of flow and transport in geological and engineering applications.
\end{abstract}

\section{Introduction}
The simulation of fluid flow and transport processes in porous materials is critical for understanding phenomena in diverse fields such as hydrogeology \cite{Bear1972}, chemical engineering \cite{Ingham2005} and materials science \cite{Torquato2002}. Advances in imaging technologies, particularly X-ray micro-computed tomography (micro-CT), have enabled the acquisition of high-resolution 3D datasets that reveal the intricate pore-scale structures of rocks and other porous materials \cite{KETCHAM2001,MA2016193}. However, the increasing resolution and size of these datasets present significant computational challenges.

Traditionally, simulations of flow and transport in porous materials have been performed using Pore Network Modelling (PNM), which simplifies the complex pore space into an interconnected network of channels and nodes \cite{Blunt2001}. While PNM offers computational efficiency and can capture key macroscopic properties, it often relies on assumptions that limit its ability to represent the detailed physics of flow and transport at the pore scale. In contrast, Direct Numerical Simulations (DNS), which solve the governing fluid flow equations directly on the pore geometry using traditional CFD methods \cite{Piller2009} or Lattice Boltzmann Method (LBM) \cite{Pan2001}, have gained increasing popularity due to advances in computational infrastructure. The availability of massively parallel supercomputers, improvements in processor speed and memory bandwidth, and scalable open-source CFD frameworks like OpenFOAM \cite{openfoam2021} now enable high-fidelity, large-scale simulations that were previously computationally prohibitive. DNS therefore provides a powerful and accurate approach to studying flow and transport phenomena in complex porous media.

Despite these advantages, DNS faces significant computational challenges. Although micro-CT imaging can now generate datasets that span several thousand voxels in each direction, DNS is typically restricted to smaller subvolumes, often limited to only a few hundred voxels per dimension, due to memory and processing limitations \cite{Giudici2023}. For larger-scale simulations, researchers frequently revert to PNM to overcome these constraints. Addressing this gap requires scalable computational frameworks and access to high-performance computing resources.

A significant challenge in applying CFD to micro-CT images lies in the generation of a computational mesh \cite{Kuhlmann31122022}. Traditional CFD approaches require meshing the pore space, where conservation equations are solved with boundary conditions applied on the pore surfaces. However, creating such a mesh for high-resolution micro-CT datasets is computationally intensive, requiring substantial memory and processing power \cite{Faieghi2019}. This challenge becomes even more pronounced for ultra-large datasets, as the process of generating complex meshes in parallel does not scale efficiently with increasing image size, often becoming a bottleneck in simulations. Recently, \cite{Wang2023} used the LBM method to perform large-scale, physically accurate modeling of proton exchange membrane fuel cells in an image that includes over 30 billion cells, using over 20,000 CPU cores. However, the LBM uses a regular lattice that does not capture the complexity of the solid surface \cite{CHANG2009940}, and LBM is somewhat difficult to extend to reactive transport modelling \cite{Liu2021}. 

To address these limitations, Immersed Boundary Methods (IBM) have emerged as a promising alternative \cite{Mittal2005}. IBM eliminates the need for body-fitted meshes by allowing simulations to be performed on simpler Cartesian grids. Various IBM techniques, such as continuous forcing \cite{PESKIN1972252}, discrete forcing with level-set \cite{CHENY20101043}, ghost-fluid \cite{MITTAL20084825} or cut-cell \cite{Trebotich2015} methods, and the penalization approach based on viscosity \cite{VINCENT2007902} or on the Darcy-Brinkman-Stokes (DBS) formulation \cite{Angot1999}, have been developed to model the presence of obstacles. For a comprehensive review of these methods, readers are referred to the work of \cite{Mittal2023}.

Recently, \cite{Trebotich2015} employed an IBM with a cut-cell method to simulate fluid flow in large binary images of porous materials. They demonstrated good scalability, simulating a synthetic image with approximately 8 billion cells across 130,000 CPU cores, yielding weak scalability at about 30,000 cells per processor. Their approach was also applied to real micro-CT images, including a CT scan of Bedford limestone (2,048x2,048x320 voxels) and a FIB-SEM scan of a shale (1,920x1,600x640 voxels). Despite these promising results, the images remain smaller than typical CT scans, and the simulations remained computationally intensive, taking between 20 and 40 seconds per time-step. This high computational cost is likely due to the explicit implementation of no-slip boundary conditions through the cut-cell method.

The DBS approach may offer a promising alternative because of its computational efficiency and simplicity. Unlike other methods, DBS does not require explicit interface localisation or complex geometric reconstructions. Furthermore, it can handle non-binary images and unresolved regions within micro-CT datasets by incorporating local permeability values, making it a more robust solution for large and complex datasets. The DBS approach has been successfully employed to simulate flow and heat transfer in micro-CT images of porous materials \cite{Maes2021}, mineral dissolution and precipitation in carbonate rocks \cite{Soulaine2017,Yang2021,Maes2022} and flow in multiscale images \cite{Soulaine2016,Carrillo2020,MAES2024113729}.

In this study, we present an open-source workflow that generates Cartesian meshes in parallel for micro-CT images and solves fluid flow and transport processes using the Darcy-Brinkman-Stokes (DBS) approach. This workflow is integrated into GeoChemFoam, an OpenFOAM-based solver (available at \href{https://github.com/GeoChemFoam}{https://github.com/GeoChemFoam}), and is accessible as a module on the UK supercomputer ARCHER2 (using \texttt{module load gcfoam}). We apply the workflow to two micro-CT images (Bentheimer sandstone and Estaillades carbonate) that includes 41 and 8 billion voxels, respectively. Our objective is to perform flow and transport simulations on the full images using up to 1,000 nodes, which represents our allocation limit on ARCHER2.  We present both strong and weak scaling results and demonstrate the ability to create a computational mesh for the entire images in under a minute, simulate flow in under an hour, and complete a full transport simulation in under 10 hours. In contrast to standard surface-conforming meshing workflows, which cannot be applied at this scale due to prohibitive memory and preprocessing demands, our approach enables efficient, high-fidelity modelling of multi-billion voxel micro-CT datasets.

\section{Method}
\subsection{Governing Equations}
The full domain $\Omega$ is made of the fluid domain $\Omega_f$ and the solid domain $\Omega_s$.
The Navier-Stokes equations for the steady laminar motion of an incompressible Newtonian fluid  $\Omega_f$ are \cite{Patankar}
\begin{align}
&\nabla\cdot\mathbf{u} = 0,\label{Eq:cont1}\\
&\nabla\cdot\left(\mathbf{u}\otimes\mathbf{u}\right) =-\nabla p +\nu\nabla^2\mathbf{u},\label{Equ:momentum1}
\end{align}
with a no-slip boundary condition at the fluid-solid interface $\Gamma$,
\begin{align}
&\mathbf{u}=0 \hspace{0.5cm} \text{at $\Gamma$},\label{Equ:bcu1}
\end{align}
where $\mathbf{u}$ (m/s) is the velocity, $p$ (m$^2$/s$^2$) is the kinematic pressure and $\nu$ (m$^2$/s) is kinematic viscosity. Our simulations are performed using a constant kinematic pressure drop $\Delta P$ between the inlet and the outlet. The permeability of the sample is defined as
\begin{equation}
    K=\frac{\nu U_DL}{\Delta P},
\end{equation}
where $U_D$ is the Darcy velocity and $L$ is the sample length. The Darcy velocity is defined as
\begin{equation}
    U_D=\frac{Q}{A},
\end{equation}
where $Q$ is the volumetric rate across the sample and $A$ is the sample cross-sectional area. The permeability is independent of the pressure drop if the Reynolds number, which characterises the relative impact of viscous force and inertia is small, i.e.
\begin{equation}
    Re=\frac{U_DL_{pore}}{\phi\nu}<<1,
\end{equation}
where $\phi$ is the sample porosity and $L_{pore}$ is the average pore size of the sample. In our simulations, we use the viscosity of water ($\nu=10^{-6}$ m$^2$/s) and the pressure drop is adjusted so that $Re=0.001$, in which case the Navier-Stokes equations can be simplified into the Stokes equations
\begin{align}
& \nabla\cdot\mathbf{u} = 0,\label{Eq:cont}\\
&0 =-\nabla p +\nu\nabla^2\mathbf{u},\label{Equ:momentum}
\end{align}

For a species transported in $\Omega_f$, the conservation can be written in terms of an advection-diffusion equation \cite{Patankar}
\begin{equation}\label{Equ:C}
 \frac{\partial C}{\partial t} + \nabla\cdot\left(C\mathbf{u}\right) -\nabla\cdot\left(D\nabla C\right)=0,
\end{equation}
where $C$ is the dimensionless concentration of the species and $D$ (m$^2$/s) is the species molecular diffusion coefficient. The P\'eclet number, which characterises the relative importance of advection and diffusion, is defined as
\begin{equation}
    Pe=\frac{U_D L_{pore}}{\phi D}.
\end{equation}
In our simulations, we use a constant concentration of $1.0$ at the inlet, and a diffusion coefficient $D=10^{-9}$ m$^2$/s, relevant for chemical species in water. With the pressure drop adjusted so that $Re=0.001$, this corresponds to $Pe=1$.

\subsection{Darcy-Brinkman-Stokes formulation}

To capture the influence of complex solid geometries, we use a volume-averaged approach. The fluid-solid interface is described in terms of the volume fraction of fluid $\varepsilon_f$ in each control volume. Mass and momentum conservation are solved in terms of the volume-averaged properties
\begin{align}
 &\overline{\mathbf{u}}=\frac{1}{V}\int_{V_f}\mathbf{u}dV,\\
 &\overline{p}_f=\frac{1}{V_f}\int_{V_f}pdV,\\
 &\overline{C}_f=\frac{1}{V_f}\int_{V_f}CdV,
\end{align}
where $V_f$ and $V$ are the volume of fluid and total volume in a control volume. Note that the velocity is scaled over $V$ (i.e., Darcy velocity) while pressure and concentration are scaled over $V_f$. The volume-averaged velocity satisfies the Darcy-Stokes-Brinkman equation over the full domain $\Omega$ \cite{Angot2017}
\begin{align}
&\nabla\cdot\overline{\mathbf{u}} = 0 \label{Eq:contDBS}\\
&0=-\nabla \overline{p}_f +\nabla\cdot\left(\frac{\nu}{\varepsilon_{f}}\nabla\overline{\mathbf{u}}\right)-\nu K_f^{-1}\overline{\mathbf{u}}\label{Eq:momentumDBS}
\end{align}
where $K_f$ is the permeability of the control volume. $\nu K_f^{-1}\overline{\mathbf{u}}$ represents the momentum exchange between the fluid and the solid phase, i.e. the Darcy resistance. This term is dominant in the solid and micro-porous phases, and vanishes in the fluid phase. It is important to note that to avoid a division by zero, $\varepsilon_f$ is assumed to be equal to a small value (i.e. 0.0001) in the solid phase.

The volume-averaged concentration satisfies the volume-averaged advection-diffusion equation \cite{Soulaine2016}
\begin{equation}\label{Equ:Cf}
 \frac{\partial \varepsilon_f\overline{C}_f}{\partial t} + \nabla\cdot\left(\overline{C}_f\overline{\mathbf{u}}\right) -\nabla\cdot\left(\varepsilon_f D^*\nabla \overline{C}_f\right)=0,
\end{equation}
where $D^*$ is the dispersion tensor within the voxel. In control volumes that are fully included in the fluid domain, the dispersion tensor is simply equal to the molecular diffusion coefficient. However, in cells that include solid or under-resolved pores, the dispersion coefficient is a function of the velocity field (hydrodynamic dispersion) and the tortuosity of the unresolved structure. For simplification, we assume
\begin{equation}
    D^*=\varepsilon_f D.
\end{equation}

The use of this method to simulate flow and transport in porous media has been extensively validated \cite{Soulaine2016b,Soulaine2016,Soulaine2017,2022-Menke,MAES2024113729,WOS:000953503400001}.
\subsection{Sample 1: Bentheimer sandstone}
\label{sec:sample1}

\begin{figure}[!t]
    \centering
\includegraphics[width=1.0\linewidth]{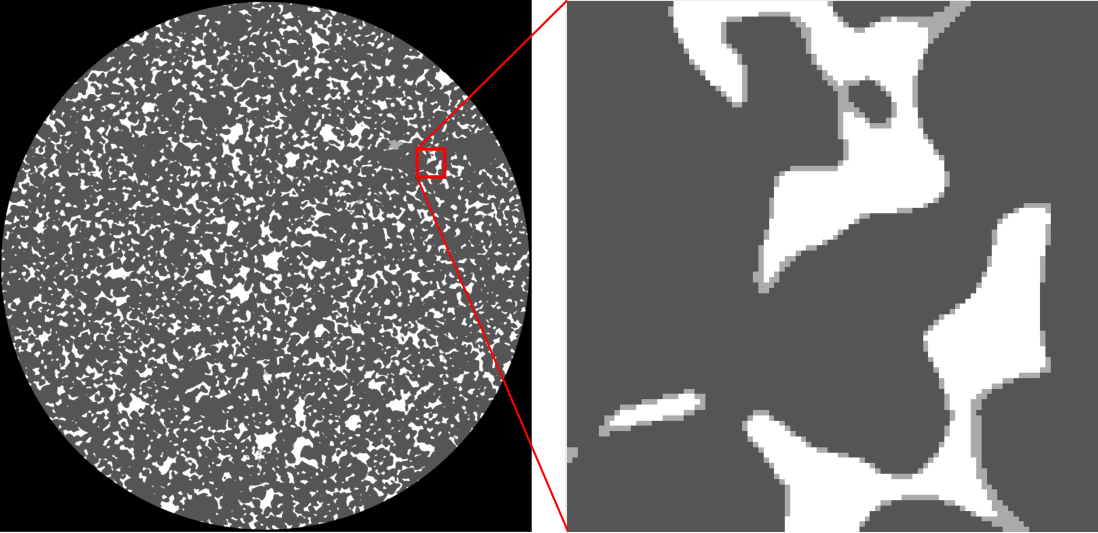}
    \caption{Micro-CT image of Bentheimer (from \cite{Bentheimer}). Black represents vitton. Greys are assumed to be solid, and white is pore. The solid surfaces are smoothed to remove the artificial roughness generated by the segmentation}
    \label{fig:BentheimerSlice}
\end{figure}

Bentheimer is a shallow marine sandstone which consists of around 95\% quartz with minor feldspar and clay and has a well-sorted grain size distribution with an average pore radius of 40 microns \cite{PEKSA2015701}. The micro-CT image has been obtained from the digital rock portal \cite{Bentheimer}. The image includes 1,950x1,950x10,800 voxels at a resolution of 6 $\mu$m,  taken using a Zeiss Versa 510 X-ray microscope. The image has been segmented with manual thresholding. The porosity, obtained from the image after the segmentation is equal to 0.18, and the permeability, measured with nitrogen using a Ruska gas permeameter \cite{PEKSA2015701} is equal to 1.64 mD. Although Bentheimer is a relatively homogeneous media for which geometrical properties such as porosity and pore-size distribution can be analysed on a much smaller domain, such a large domain is useful to characterise a Representative Elementary Volume (REV) for flow properties such as absolute and relative permeability \cite{Jackson2020}.

The segmented micro-CT image includes 4 labels: viton, resolved solid, unresolved pores and resolved pores. In our study, we assume that unresolved is solid (i.e. $\varepsilon_f=0.0001$), but the solid surfaces are then smoothed using a Lagrangian smoother to eliminate the artificial roughness generated by the segmentation (Fig. \ref{fig:BentheimerSlice}). The smoothing introduces porosity values between 0.0001 and 1.0, and the local permeability for every voxel needs to be modelled. For this, we use a Kozeny-Carman relationship \cite{Soulaine2016b}
\begin{equation}\label{Equ:KK}
    K_f^{-1}=A_{KC} \frac{\left(1-\varepsilon_f\right)^2}{\varepsilon_f^3},
\end{equation}
and $A_{KC}$ is the Kozeny-Carman constant. The value of $A_{KC}$ may impact the flow field and the prediction of the rock sample permeability. It is a subject of active research \cite{Soulaine2016b,Krotkiewski_Ligaarden_Lie_Schmid_2011}, but is beyond the scope of our work. In this study, we use:
\begin{equation}
    A_{KC}=\frac{180}{h^2}
\end{equation}
where h is the image resolution, i.e. 6 $\mu$m. This formula yields a low permeability of $2\times 10^{-25}$ mD in the solid phase.

\subsection{Sample 2: Estaillades Carbonate}
\label{sec:sample2}
    
Estaillades is a limestone made mostly of calcite (>97\%) with a minor quartz component. It shows a bimodal pore-size distribution with micro-porosity within the grains. The macro-pores have an average radius of approximately 10 microns, while the micro-pores with average radius of approximately 0.4 microns \cite{2022-Menke}. The micro-CT image has been obtained from the BGS repository \cite{menke2018_estaillades_limestone}. The image includes 1,202x1,236x6,000 voxels at a resolution of 3.9676 $\mu$m,  taken using a Zeiss XRM-510 micro-CT. The Weka3D machine learning segmentation algorithm in Fiji \cite{schindelin2012fiji} was used to segment the macro-pores. Images of the rock filled with doped brine were then used to identify the solid grains and unconnected micro-porosity. The pore space, unconnected micro-porosity, and solid grains were then masked, and the remaining greyscale values were used to segment the connected micro-porous grains into 12 micro-porous phases (i.e. labels) based on porosity \cite{2022-Menke}. Due to segmentation problems at the solid/viton interface, we selected a subvolume of size 1,144×1,144×6,000 to avoid these issues while still retaining a sufficiently large problem size for scalability and performance evaluation. Fig. \ref{fig:EstailladesSlice} shows a slice of the segmented image. The sample porosity, obtained from the image after the segmentation, and permeability, measured with brine using a Keller PD-33X differential pressure transducer \cite{2022-Menke} are equal to 0.34 and 25mD, respectively.

\begin{minipage}{0.55\linewidth}
		\centering
		\label{figu:re}
		\includegraphics[width=0.95\textwidth]{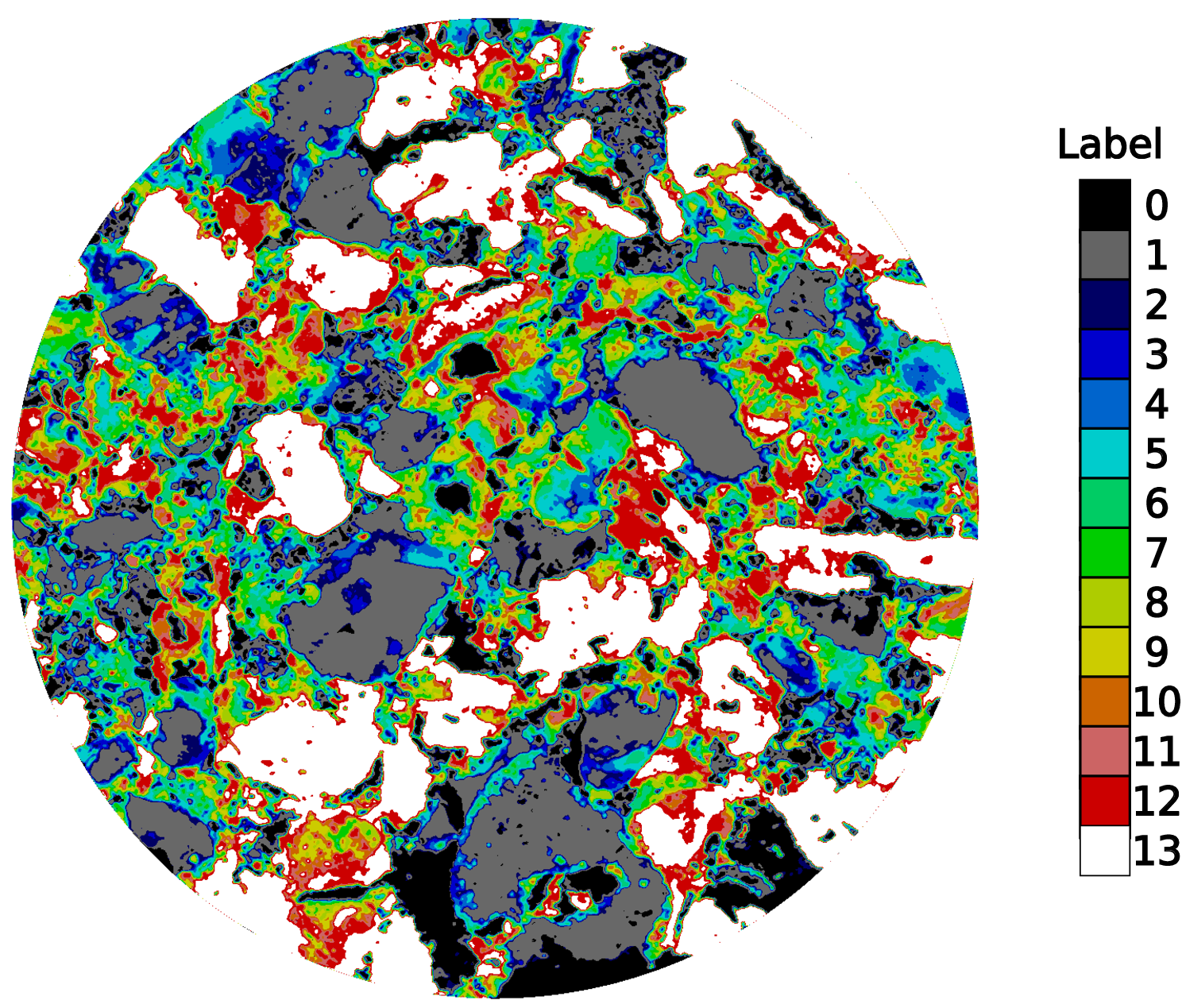}
        		\captionof{figure}[Caption1]{Middle slice of the Estaillades micro-CT image segmented into 14 labels  \label{fig:EstailladesSlice}}
	\end{minipage}\hfill
	\begin{minipage}{0.4\linewidth}
		\label{tab:le}
		\centering
            \renewcommand{\arraystretch}{1.5}
		\resizebox{\textwidth}{!}{%
		\begin{tabular}{c|ccc}
        Label & Fraction & $\phi$ & $K_f$ (mD) \\
        \hline
        0 & 0.100 & 1.0 & $\infty$ \\[-1ex]
        1 & 0.172 & 0.57 & 7.57 \\[-1ex]
        2 & 0.046 & 0.52 & 7.00 \\[-1ex]
        3 & 0.046 & 0.47 & 4.85 \\[-1ex]
        4 & 0.049 & 0.42 & 3.28 \\[-1ex]
        5 & 0.085 & 0.36 & 2.15 \\[-1ex]
        6 & 0.072 & 0.27 & 0.817 \\[-1ex]
        7 & 0.041 & 0.22 & 0.465 \\[-1ex]
        8 & 0.037 & 0.18 & 0.247 \\[-1ex]
        9 & 0.032 & 0.15 & 0.119 \\[-1ex]
        10 & 0.028 & 0.12 & 0.0502 \\[-1ex]
        11 & 0.024 & 0.09 & 0.0175 \\[-1ex]
        12 & 0.077 & 0.07 & 0.00482 \\[-1ex]
        13 & 0.192 & 0.0001 & 0.000001 \\
    \end{tabular}}
    		\captionof{table}{Total volume fraction, porosity and permeability of each segmentation phase (i.e., label) for the Estaillades micro-CT images. \label{tab:EstailladesPoroPerm}}
            \vspace{0.2cm}
	\end{minipage}
    
The porosity and permeability of each segmented phases are presented in Table \ref{tab:EstailladesPoroPerm}. To estimate the micro-porous permeability, a nano-CT image (using Zeiss Ultra Nano-CT) was obtained from a subsample of 65 microns diameter, drilled within a region included in label 4 ($\phi=0.42$). The image was then analyzed for a range of properties (e.g., pore size and grain size distribution) and the information was used as input into an object-based pore network generator, on which permeability fields were simulated for a range of porosities \cite{2022-Menke}, creating a synthetic porosity-permeability relationship that was used for each label (Table \ref{tab:EstailladesPoroPerm}). For the pores and solid phase, we assume $K_f^{-1}=0$ and $K_f=10^{-6}$ mD, respectively.

\subsection{Parallel meshing}
The first step of CFD requires the building of a computational mesh. In OpenFOAM, the mesh is defined within the \texttt{constant/polyMesh} 
directory of a case file \cite{openfoam2021}. This directory contains several files that describe the geometry, connectivity, and properties of the computational mesh. Firstly, the \texttt{points} file lists all vertices (i.e. 3D coordinates) of the mesh. These are the points that form the corners of computational grid blocks. These points are shared among neighbouring cells and define the geometry. For hexahedral grid blocks, each cell is formed of 8 points, and each point 
can belong to up to 8 grid blocks. Then, the \texttt{face} file describes the faces of the mesh in terms of vertex indices. For hexahedral grid blocks, a face is made of 4 points. Faces can be categorised between internal faces, i.e. which belong to two neighbouring cells, and boundary faces, which belong to a grid block and a boundary of the domain.

In OpenFOAM, the internal faces are indexed first before all boundary faces. The \texttt{owner} and \texttt{neighbour} files map faces to cells: \texttt{owner} specifies the cell that "owns" each face, while \texttt{neighbour} specifies the adjacent cell for internal faces only. Boundary faces, which do not have adjacent cells, appear only in the owner file and are further detailed in the \texttt{boundary file}, which specifies boundary conditions and regions. The strict ordering of internal faces first ensures alignment with the neighbour file, as it relies on the sequential order to define connectivity. It is important to note that the representation of a single mesh is not unique, as it depends on the ordering of points, faces and cells, which is algorithm-defined.

In an OpenFOAM parallel (MPI) mesh, grid blocks are divided into subdomains, each described in its own folder \texttt{processorJ}, where $J$ is the processor's rank. Each \texttt{processor} folder contains its own \texttt{constant/polyMesh} subdirectory with local mesh files. As a result, the MPI mesh can be viewed as a collection of sub-meshes, one for each processor. The \texttt{points} file defines the vertices of the processor mesh independently. The \texttt{faces}, \texttt{owner}, and \texttt{neighbour} files describe the sub-mesh faces in terms of internal and boundary faces. Faces shared between grid blocks belonging to different processors are internal faces in the global full mesh but appear as boundary faces in the processor sub-meshes. These additional boundaries are identified as 
\texttt{procBoundary} in the \texttt{boundary} files. Moreover, four additional files—\texttt{cellProcAddressing}, \texttt{pointProcAddressing}, \texttt{faceProcAddressing}, and \texttt{boundaryProcAddressing}—are used to map indices in the processor sub-meshes to those in the global full mesh. These files enable efficient communication and data exchange between subdomains during parallel computations.

In an unstructured mesh, there is no direct relationship between point indices and their coordinates, making it challenging to construct the mesh quickly and efficiently, especially in parallel. However, our numerical approach allows for the use of a Cartesian grid, where point indices are directly tied to coordinates. This capability is particularly advantageous for parallel processing.

\begin{figure}[!t]
    \centering
\includegraphics[width=1.0\linewidth]{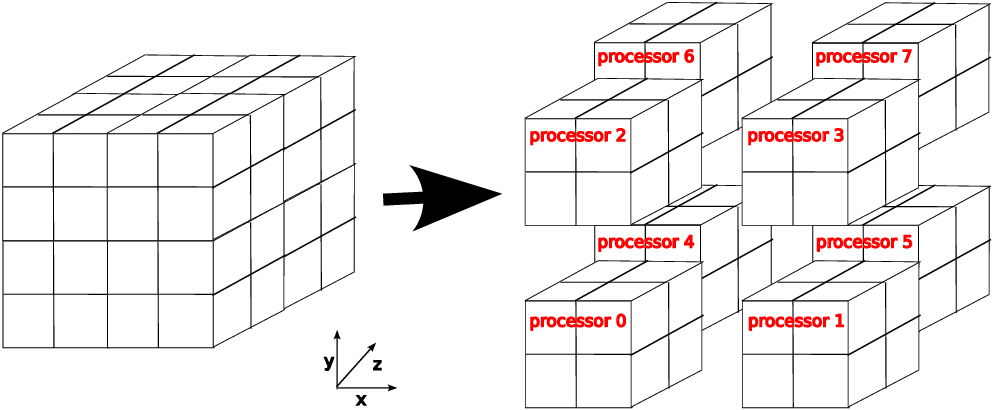}
    \caption{ Illustration of the regular Cartesian mesh decomposition. The full domain (left) is split into equally sized subdomains (right), each assigned to a processor. Processor indices increase first along the x-direction, then y, and finally z, enabling a predictable mapping of subdomains.}
    \label{fig:grid}
\end{figure}

In our approach, the processor sub-meshes are defined regularly, as shown in Figure \ref{fig:grid}. Each processor can then construct its local mesh independently. In addition, it is straightforward to calculate the global index of a mesh point based on its local index in its processor, and on the processor's rank. Therefore, each processor can independently generate its local mesh files, including \texttt{cellProcAddressing}, \texttt{pointProcAddressing}, \texttt{faceProcAddressing}, and \texttt{boundaryProcAddressing}, enabling high scalability.

To determine the mapping between global and local indices, we consider a micro-CT image with dimensions \(N_X \times N_Y \times N_Z\), and a parallel decomposition defined by \(N_{PX} \times N_{PY} \times N_{PZ}\) processors in the \(x\), \(y\), and \(z\) directions, respectively. For each direction \(J \in \{X, Y, Z\}\), if \(N_J\) is evenly divisible by \(N_{PJ}\), then the number of cells per processor in that direction, \(N_{JPP}\), is constant and equal to \(Q_J = N_J / N_{PJ}\). 

In cases where the number of processors $N_{PJ}$ does not evenly divide the number of cells $N_J$, we define the remainder of the division as $r$, such that 
$N_J=Q_J\times N_{JP}+r$. In this situation, the first $r$ processors along direction $J$ are assigned sub-blocks that are one line of cells larger ($Q_J$+1) than those of the remaining processors ($Q_J$). This leads to a non-uniform distribution of data. The indexing scheme nevertheless allows each processor to compute the global indices of its local cells directly from its rank and local indices, without requiring communication. 

The final step in the mesh generation process involves defining the local porosity and permeability fields. In our approach, these are represented by two \texttt{volScalarField} objects: \texttt{eps}, corresponding to the local porosity, and \texttt{Kinv}, representing the inverse of the local permeability. In OpenFOAM, a \texttt{volScalarField} is a field defined over the cell centers (volume elements) of the computational mesh, used to store scalar quantities such as pressure, temperature, or, in this case, porosity and permeability \cite{openfoam2021}. Both fields are initialised by reading their values from the so-called \texttt{0} directory of the test case at the beginning of the simulation. In parallel executions, \texttt{eps} and \texttt{Kinv} are decomposed such that each processor initializes its internal cells and boundary conditions—including the \texttt{procBoundary}—by reading data from the corresponding \texttt{processorJ/0} directory. Each input file contains a list of values corresponding to the field values of the internal cells, ordered according to the internal cell numbering, as well as values for the \texttt{procBoundary} and any global boundaries assigned to that processor.

By employing a Cartesian mesh that precisely matches the dimensions of the micro-CT image, along with a regular parallel decomposition, we ensure a direct correspondence between the internal numbering of both internal and boundary cells within each processor and their respective indices in the micro-CT image. As a result, each processor can independently construct its own \texttt{eps} and \texttt{Kinv} field files, which supports a highly scalable parallel implementation.

Each processor opens the micro-CT image, which is stored in a \texttt{.raw} format with dimensions \(N_X\), \(N_Y\), and \(N_Z\). This format corresponds to \(N_Z\) layers, where each layer is a two-dimensional array of size \(N_X \times N_Y\). To minimize memory usage and I/O overhead, the image is read layer by layer, and a processor accesses a given layer only if it contains local internal or boundary cells. The global indices of these cells are then mapped to the corresponding local indices within the processor and stored in a three-dimensional array of size \(N_{XPP} \times N_{YPP} \times N_{ZPP}\), which reflects the number of cells assigned to each processor. Each voxel label is subsequently translated into corresponding values of \texttt{eps} and \texttt{Kinv}. For the Bentheimer dataset, labels 1, 2, and 3 represent solid regions (\(\varepsilon_f = 0.0001\)), while label 4 represents pore space (\(\varepsilon_f = 1.0\)). The image is smoothed, and the permeability field is computed using Equation~(\ref{Equ:KK}). For the Estaillades sample, voxel labels are mapped to porosity \(\varepsilon_f\) and permeability \(K\) using the values provided in Table~\ref{tab:le}.


\section{Results}

In this section, we evaluate the efficiency of our workflow in terms of mesh generation and solver performance, both through direct comparison with the standard OpenFOAM approach and through strong and weak scaling studies. The benchmarks use cropped subvolumes from the Bentheimer and Estaillades samples described previously. Since the goal is to simulate the full images using no more than 1,000 nodes, the selected configurations are constrained to a maximum of approximately 320,826 cells per core for Bentheimer and 61,347 cells per core for Estaillades. 

While the mesh generation algorithm is deterministic, its runtime on ARCHER2 is subject to significant variability caused by file system I/O performance. We observed that I/O overheads could account for anywhere between 10\% and 99\% of the total mesh generation time, primarily due to contention on the shared Lustre file system~\cite{archer2docs}. For example, running the mesh generation step for the full Bentheimer sample (1,950 × 1,950 × 10,800 voxels) five times on 65,536 cores (512 nodes) resulted in total runtimes of 545, 52, 221, 612, and 3,720 seconds. However, in all cases, the actual computational time was between 25 and 26 seconds, with the remainder dominated by variability in file system performance. To ensure fair and consistent comparisons, we report only the computational mesh generation time, excluding I/O.

Flow simulations are performed using the standard Semi-Implicit Method for Pressure-Linked Equations (SIMPLE) with under-relaxation~\cite{Patankar}. Each non-linear iteration involves solving a sequence of linear systems. Two linear solvers available in OpenFOAM are considered: the Geometric-Algebraic Multi-Grid (GAMG) solver and the Preconditioned Bi-Conjugate Gradient Stabilised (PBiCGStab) solver~\cite{openfoam2021}. Full implementation details for the solvers (\texttt{simpleDBSFOAM}), test cases, and parameter settings are available in the GeoChemFoam project repository at \href{https://github.com/GeoChemFoam}{https://github.com/GeoChemFoam}.

For the comparison with standard approach and strong scaling tests, simulations are run to steady-state, defined by a pressure residual below $10^{-4}$. Within each SIMPLE iteration, the linear systems are solved to a relative tolerance of 0.1. The transport solver is executed over 100 time steps of fixed duration $\Delta t = 0.001$ seconds, corresponding to an average CFL number of 0.05. For each time step, the linearised equation is solved using PBiCGStab until the residual drops below $10^{-9}$. 

In the strong scaling test, since the amount of computations required remains the same, the strong scaling parallel efficiency $\mbox{\it{Eff}}_s$ is defined as
\begin{equation}
\mbox{\it{Eff}}_s =\frac{T_{Ref}}{T}\frac{N_{P,Ref}}{N_P},
\end{equation}
where $T$ is the computational time for the simulation and $N_P$ is the number of processors (cores) used, while $T_{Ref}$ and $N_{P,Ref}$ are the values for the baseline case. Ideally, the baseline case runs on 1 core or, due resource constraints, the lowest core count possible. For both the following strong and weak scaling studies, a single node (128 cores) was used as the baseline case due to memory constraints.  This choice reflects the practical limitations of the target large-scale Bentheimer simulation, which requires the memory capacity of at least 100 nodes, and also aligns with realistic deployment scenarios for large-scale OpenFOAM cases. The corresponding memory requirement is estimated at roughly 300 bytes per cell, comprising about 200 bytes per cell for mesh topology and geometric data (points, faces, owner–neighbour connectivity, and geometric fields) and around 100 bytes per cell for field variables and solver data. These figures are consistent with previously reported OpenFOAM memory and scalability characteristics \cite{Weller1998,Jasak2009, Higuera2014, PRACE2013}. Using this full node for parallel efficiency baselines allowed us to evaluate parallel performance within the limits of available resources and time.

In the weak scaling tests, where the domain size increases with core count, the solver convergence behaviour can be affected due to changes in geometric complexity. To isolate parallel efficiency from convergence variability, we fix the number of iterations per time step: one iteration for GAMG (sufficient for approximate convergence at a relative tolerance of ~0.1), and 30 iterations for PBiCGStab (to achieve comparable accuracy). The transport solver is similarly executed with 5 fixed iterations per time step using PBiCGStab. This fixed-iteration strategy ensures a constant computational workload per time step, allowing us to assess the impact of parallelism independently of algorithmic convergence. The weak scaling parallel efficiency $\mbox{\it{Eff}}_w$ is defined as
\begin{equation}
\mbox{\it{Eff}}_w =\frac{T_{Ref}}{T}.
\end{equation}

\subsection{Comparison with standard approach}

To evaluate the practical benefits of our proposed workflow, we compared it with the conventional mesh-based approach typically used in OpenFOAM, which involves generating a conformal mesh. A clear advantage of our method is that it enables direct integration of under-resolved porosity, as in the Estaillades sample. However, it also provides substantial benefits for simpler geometries like the Bentheimer sample. In this subsection, we consider a $600^3$ subvolume of Bentheimer extracted in the middle of the sample. The mesh is generated, using the standard OpenFOAM approach, and our novel approach, then the flow is solved until convergence using the GAMG linear solver. Table~\ref{tab:comparison} summarizes the timing results for both workflows, tested across increasing node and core counts, ranging from 1 node (128 cores) to 5 nodes (640 cores).

\begin{table}[!t]
\centering
\caption{Comparison of simulation timings between the standard and new workflows. The standard approach uses \texttt{blockMesh}, \texttt{decomposePar}, and \texttt{snappyHexMesh} for mesh generation. The new approach constructs the mesh directly in parallel using a Cartesian grid. Flow times reflect single-phase steady-state solver runtimes.}
\label{tab:comparison}
\begin{tabular}{cccccc|cc}
\toprule
\textbf{Cores} & \textbf{Nodes} & \multicolumn{4}{c|}{\textbf{Standard approach}} & \multicolumn{2}{c}{\textbf{New approach}} \\
& & \textbf{blockMesh (s)} & \textbf{decomposePar (s)} & \textbf{snappyHexMesh (s)} & \textbf{Flow (s)} & \textbf{createMesh (s)} & \textbf{Flow (s)} \\
\midrule
128 & 1   & 23 & 67  & 338 & 1,210 & 54  & 6,710 \\
256 & 2   & 23 & 77  & 273 & 525  & 28  & 3,021 \\
384 & 3   & 23 & 96 & 257 & 328  & 17  & 1,973 \\
512 & 4   & 23 & 105 & 251 & 232  & 13  & 1,426 \\
640 & 5   & 23 & 116 & 237 & 175  & 10  & 1,131 \\
\bottomrule
\end{tabular}
\end{table}

The standard workflow involves a multistep mesh generation process: \texttt{blockMesh} defines a coarse background mesh, \texttt{decomposePar} partitions the domain for parallel execution, and \texttt{snappyHexMesh} conforms the mesh to the geometry. The initial \texttt{blockMesh} step runs on a single core and has substantial memory requirements, which prevent the direct generation of a $600^3$ background mesh. To avoid this limitation, a coarser mesh with a $150^3$ block structure is first created, then partitioned using \texttt{decomposePar}, and finally refined by a factor of four during the parallel execution of \texttt{snappyHexMesh}, effectively producing a $600^3$ background mesh. The \texttt{snappyHexMesh} utility uses an STL surface representation of the pore geometry to remove grid cells within the solid phase, then iteratively refines and morphs the mesh, snapping cell faces to the interface to accurately capture the solid–fluid boundary~\cite{openfoam2021}.

Despite this refinement, the standard workflow exhibits several scalability limitations. The \texttt{decomposePar} utility, although intended to support parallelism, operates in serial and becomes slower as the number of partitions increases. This bottleneck becomes particularly problematic in large-scale simulations. Additionally, \texttt{snappyHexMesh} includes several stages that are only partially parallelised and do not scale well with domain size or processor count. As a result, mesh generation remains time-consuming and inefficient for large domains, and becomes computationally infeasible when the domain's mesh requires more memory than is available on a single shared node.

In contrast, our workflow constructs the mesh directly in parallel using a structured Cartesian grid that aligns with the voxelised input image. This eliminates all serial preprocessing steps. Each core independently builds its portion of the mesh without inter-process communication. Consequently, mesh generation scales efficiently with core count, decreasing from 54 seconds at 128 cores to just 10 seconds at 640 cores. This scaling behaviour is further analysed in Sections~\ref{sec:strongScal} and \ref{sec:weakScal}.

\begin{figure}[!t]
\centering
\includegraphics[width=\linewidth]{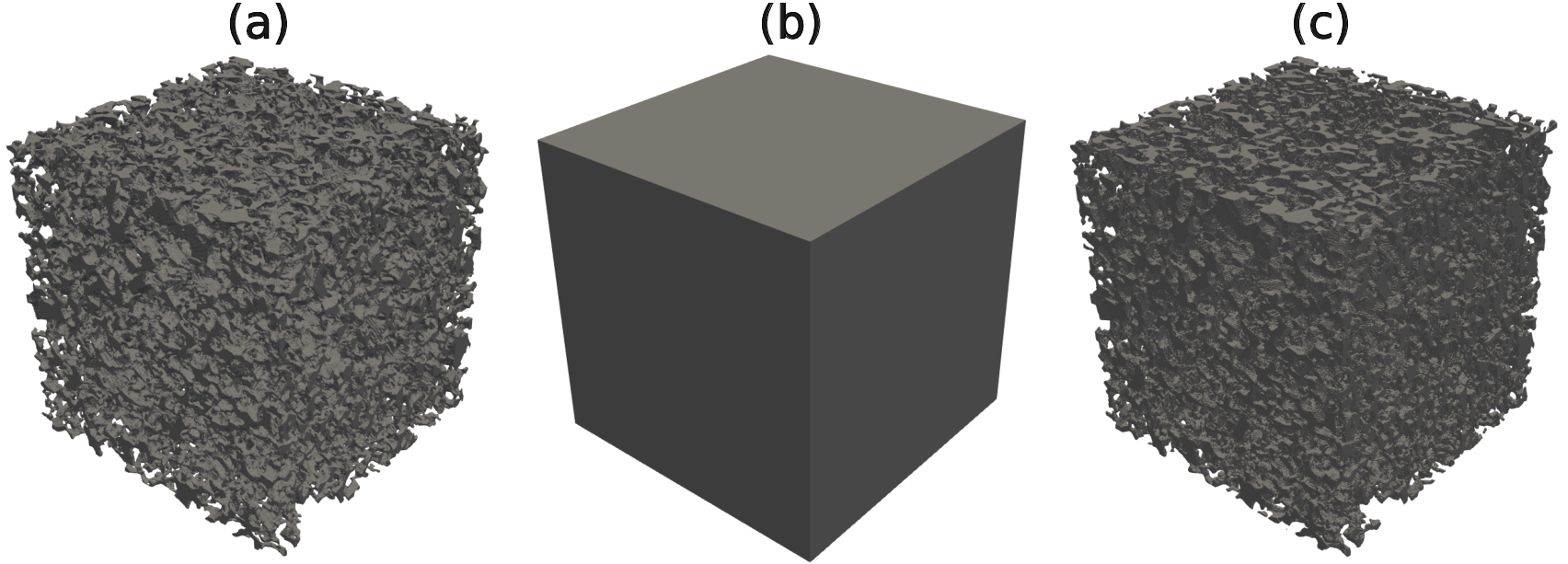}
\caption{Computational domains: (a) Standard approach (40M grid blocks); (b) New approach (225M grid blocks); (c) Fluid and interface cells in new approach (53M grid blocks).}
\label{fig:BentheimerDomains}
\end{figure}

Although flow solver times are consistently longer in our approach, this is an expected consequence of retaining the entire voxel domain, including both the solid and the fluid phases. In the standard workflow, the mesh includes only the fluid region, approximately 40 million cells in our case (Fig.~\ref{fig:BentheimerDomains}a). In contrast, our approach solves for the entire $600^3 = 216$ million cells, about 5.4 times more (Fig.~\ref{fig:BentheimerDomains}b).

The increase in solver time reflects this difference. Across all core counts, the flow time ratio between the two workflows ranges from 5 to 6, which is consistent with the increase in cell count. Importantly, while the solver time per core is higher, our approach continues to scale effectively with added resources. In contrast, the standard workflow shows limited scalability due to serial mesh generation stages.

The solver convergence behaviour is compared in Table~\ref{tab:comparison2}. The number of non-linear iterations is nearly constant across core counts for both methods, indicating that domain decomposition does not significantly impact convergence. Interestingly, the new approach consistently requires around 10\% fewer non-linear iterations, likely due to the use of a Cartesian mesh, which avoids cell skewness and ensures uniform spacing. However, the number of linear iterations per non-linear step is higher, which is expected given the larger number of cells and unknowns in the system. Despite these differences, the computed permeability is identical in all cases, validating the numerical accuracy of the new method.

\begin{table}[!t]
\centering
\caption{Comparison of solver convergence between standard and new workflows}
\label{tab:comparison2}
\begin{tabular}{cccccccc}
\toprule
\textbf{Cores} & \textbf{Nodes} & \multicolumn{3}{c}{\textbf{Standard approach}} & \multicolumn{3}{c}{\textbf{New approach}} \\
& & \textbf{Non-linear it.} & \textbf{Linear it.} & \textbf{Perm. (m$^2$)} & \textbf{Non-linear it.} & \textbf{Linear it.} & \textbf{Perm. (m$^2$)} \\
\midrule
128 & 1 & 972 & 1,641 & $2.4 \times 10^{-12}$ & 864 & 1,922 & $2.4 \times 10^{-12}$ \\
256 & 2 & 972 & 1,558 & $2.4 \times 10^{-12}$ & 868 & 1,987 & $2.4 \times 10^{-12}$ \\
384 & 3 & 972 & 1,575 & $2.4 \times 10^{-12}$ & 873 & 1,987 & $2.4 \times 10^{-12}$ \\
512 & 4 & 972 & 1,572 & $2.4 \times 10^{-12}$ & 872 & 1,985 & $2.4 \times 10^{-12}$ \\
650 & 5 & 971 & 1,635 & $2.4 \times 10^{-12}$ & 871 & 1,979 & $2.4 \times 10^{-12}$ \\
\bottomrule
\end{tabular}
\end{table}

\begin{figure}[!t]
\centering
\includegraphics[width=0.9\linewidth]{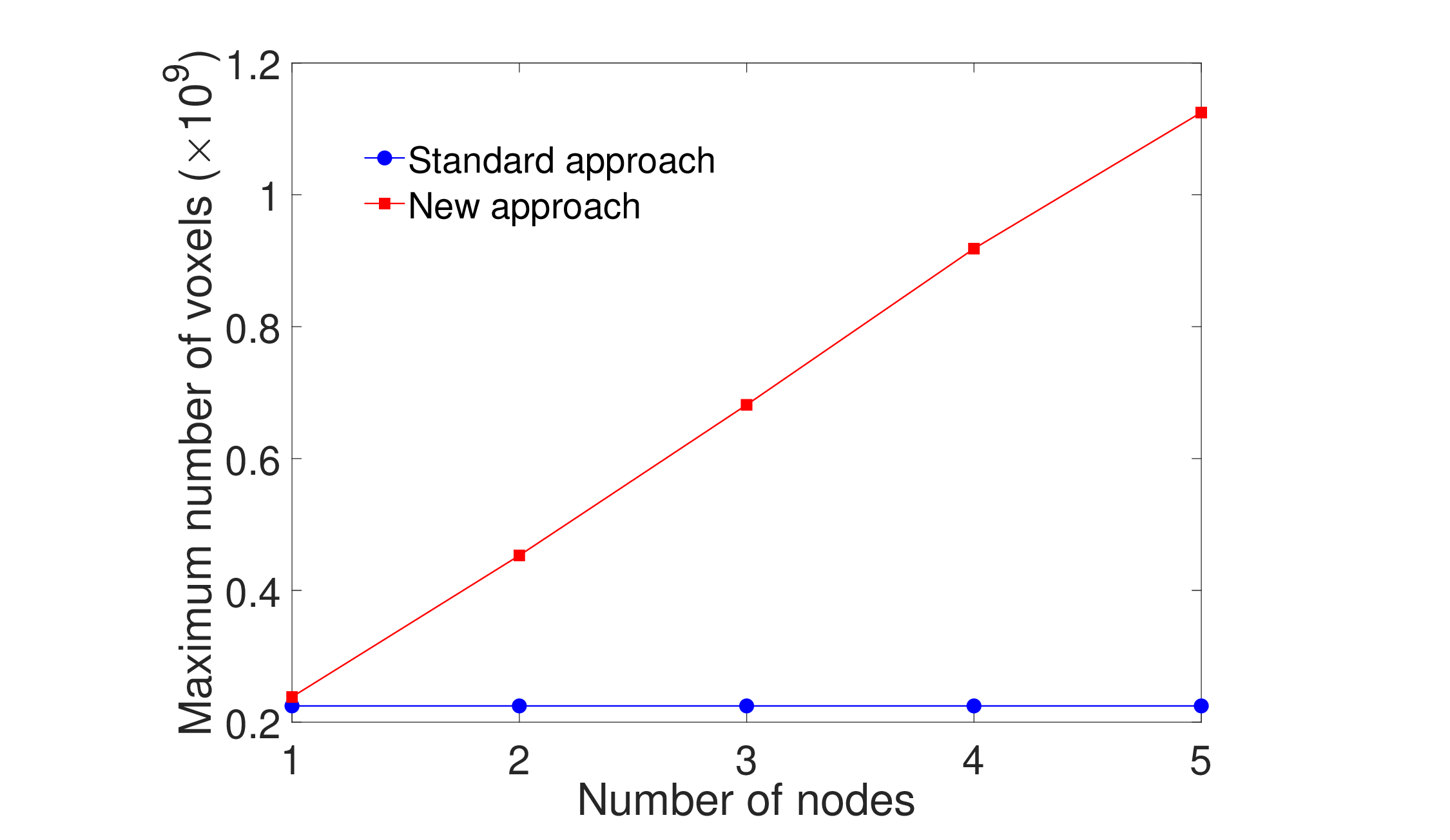}
\caption{Maximum image size (in billions of voxels) that can be processed as a function of node count for the standard and new workflows. The standard method is limited to approximately 225 million voxels. The new workflow scales linearly with node count.}
\label{fig:max_domain}
\end{figure}

To better illustrate the scaling limitations of the standard workflow, we evaluated the maximum domain size that can be meshed and simulated as a function of the number of nodes. Figure~\ref{fig:max_domain} shows the largest voxel domain processed successfully for each method. For the standard workflow, we observed a hard upper limit of approximately 225 million voxels ($608^3$), beyond which \texttt{snappyHexMesh} failed due to memory constraints. This limitation was independent of node count and arises from the need to maintain a global view of the mesh during processing, which concentrates memory usage on a single core.

In contrast, our workflow exhibits nearly linear growth in maximum processable domain size with node count. Using five nodes, we successfully handled a domain exceeding 1 billion voxels ($1040^3$). This confirms that mesh creation is fully distributed, with local memory requirements only and no global mesh assembly required. This scalability makes our approach ideally suited for simulations of large rock samples, directly achieving our goal of enabling billion-voxel image computations.

Although the solver remains slower due to the larger mesh size, several strategies can help improve performance. First, increasing the number of cores yields significant speed-ups, particularly during mesh construction, due to strong scalability. Second, the solid region can be removed after meshing using utilities such as \texttt{subSetMesh}, retaining only the fluid and interface cells (Fig.~\ref{fig:BentheimerDomains}c). However, this post-processing step requires substantial memory and scales poorly, so we avoid it in the current workflow. In future work, we plan to implement local mesh adaptation (e.g., refinement and coarsening) to reduce the number of solid-phase cells while maintaining high resolution in fluid regions.

In summary, while the standard approach yields a smaller mesh and faster solver times, it suffers from poor scalability and memory limitations. Our proposed workflow eliminates these constraints through fully parallel mesh generation and distributed memory use. The method opens the door to high-resolution simulations of very large domains, well beyond the limits of current meshing tools.

\begin{table}[!t]
\centering
\caption{Strong scaling results for the Bentheimer sub-domain ($1,950 \times 1,950 \times 30$ cells).  Parallel efficiency is relative to the 1 node configuration.}
\label{tab:strong_scal_bentheimer}
\begin{tabular}{ccccccccccccc}
\toprule
\textbf{Cores} & \textbf{Nodes} & \textbf{Cells per core} & \multicolumn{2}{c}{\textbf{Mesh Gen.}} & \multicolumn{2}{c}{\textbf{Flow (GAMG)}} & \multicolumn{2}{c}{\textbf{Flow (PBiCGStab)}} & \multicolumn{2}{c}{\textbf{Transport}} \\
& & & \textbf{Time (s)} & \textbf{Eff.} & \textbf{SPT} & \textbf{Eff.} & \textbf{SPT} & \textbf{Eff.} & \textbf{SPT} & \textbf{Eff.} \\
\midrule
        128 & 1 & 1,485,352 & 49.8  & 1.00  & 7.50  & 1.00  & 11.9   & 1.00  & 3.33  & 1.0  \\
     256 & 2 & 742,676 & 24.8  & 1.00  & 3.31  & 1.13   & 5.82  & 1.02  & 1.56 & 1.07  \\
     512 & 4 & 371,338 & 12.6  & 0.99  & 1.58  & 1.19   & 2.96  & 1.01  & 0.77 & 1.08 \\
     1,024 & 8 & 185,669 & 6.36  & 0.98  & 0.77  & 1.22   & 1.42  & 1.05  & 0.37 & 1.12  \\
     2,048 & 16 & 92,835 & 3.24  & 0.96  & 0.32  & 1.46  & 0.57   & 1.30  & 0.15 & 1.39  \\
    4,096 & 32 & 46,418 & 1.67  & 0.93  & 0.16  & 1.46  & 0.23   & 1.62  & 0.064 & 1.63  \\
    8,192 & 64 & 23,209 & 0.89  & 0.87  & 0.071  & 1.65  & 0.082   & 2.27  & 0.022 & 2.37 \\
    16,384 & 128 & 11,605 & 0.44  & 0.88  & 0.039  & 1.50  & 0.043   & 2.16  & 0.0088 & 2.96 \\
    32,768 & 256 & 5,802 & 0.22  & 0.88  & 0.028  & 1.05  & 0.03   & 1.55  & 0.0049 & 2.65 \\

\bottomrule
\end{tabular}
\end{table}

\subsection{Strong Scaling Analysis}
\label{sec:strongScal}
\begin{table}[!t]
\centering
\caption{Strong scaling results for the Estaillades sub-domain ($1,144 \times 1,144 \times 150$ cells). Parallel efficiency is relative to the 1 node configuration.}
\label{tab:strong_scal_estaillades}
\begin{tabular}{ccccccccccccc}
\toprule
\textbf{Cores} & \textbf{Nodes} &  \textbf{Cells per core} & \multicolumn{2}{c}{\textbf{Mesh Gen.}} & \multicolumn{2}{c}{\textbf{Flow (GAMG)}} & \multicolumn{2}{c}{\textbf{Flow (PBiCGStab)}} & \multicolumn{2}{c}{\textbf{Transport}} \\
& & & \textbf{Time (s)} & \textbf{Eff.} & \textbf{SPT} & \textbf{Eff.} & \textbf{SPT} & \textbf{Eff.} & \textbf{SPT} & \textbf{Eff.} \\
\midrule
128 & 1 & 1,533,675 & 59.0  & 1.00  & 6.50 & 1.00  & 22.7   & 1.00  & 9.20  & 1.00  \\
     256 & 2 & 766,838 & 29.4  & 1.00  & 3.24  & 1.00  & 11.5   & 0.99  & 4.17  & 1.10  \\
     512 & 4 & 383,418 & 14.9  & 0.99  & 1.62  & 1.00  & 5.60   & 1.01  & 2.12  & 1.08  \\
    1,024 & 8 & 191,710 & 7.57  & 0.97  & 0.81  & 1.00  & 2.53   & 1.12  & 1.01 & 1.14  \\
    2,048 & 16 & 95,855 & 3.80  & 0.97  & 0.36  & 1.13  & 1.07   & 1.33  & 0.44 & 1.31  \\
    4,096 & 32 & 47,928 & 1.90  & 0.97  & 0.16  & 1.27  & 0.40   & 1.77  & 0.18  & 1.60  \\
    8,192 & 64 & 23,964 & 0.97  & 0.95  & 0.078  & 1.30  & 0.13   & 2.73  & 0.053  & 2.71  \\
    16,384 &128 & 11.982 & 0.49  & 0.94  & 0.048  & 0.072  & 1.66   & 2.46  & 0.026  & 2.76  \\
    32,768 & 256 & 5,991 & 0.25  & 0.92  & 0.044  & 0.054  & 1.66   & 1.64  & 0.018  & 2.00  \\
\bottomrule
\end{tabular}
\end{table}

\begin{figure}[!t]
\centering
\includegraphics[width=\linewidth]{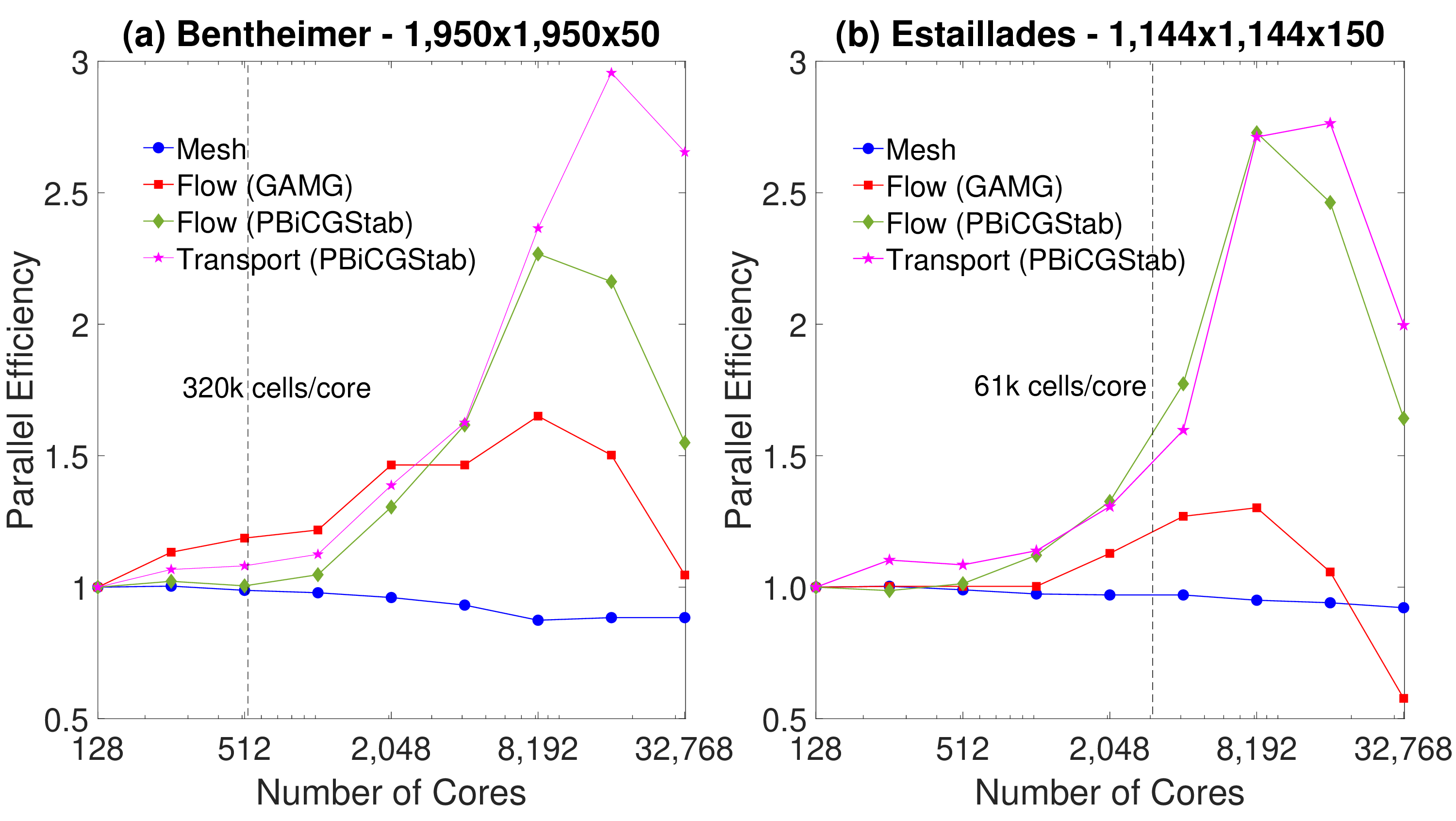}
\caption{Strong scaling results for the Bentheimer ($1,950\times 1,950 \times 50)$ and the Estaillades ($1,144 \times 1,144 \times 150$ cells) subdomains. Parallel efficiency is relative to the 1-node configuration}
\label{fig:StrongScal}
\end{figure}

To evaluate strong scalability, we consider the largest subvolume in the \( z \)-direction (i.e., number of slices) that can be processed on a single core due to memory constraints. For the Bentheimer domain, this corresponds  to \(1,950 \times 1,950 \times 50\); for Estaillades, it is \(1,144 \times 1,144 \times 150\). The strong scaling tests involve increasing the number of cores while maintaining a constant problem size (i.e., 190M cells for Bentheimer, 196M for Estaillades, ). As the number of cores increases, the number of cells per core decreases, allowing us to assess the efficiency with which each solver leverages additional computational resources under fixed workload conditions.

Tables~\ref{tab:strong_scal_bentheimer} and \ref{tab:strong_scal_estaillades} present the execution times and corresponding parallel efficiencies for each configuration and for each stage of the simulation. Results are shown for mesh generation (in total seconds) and for the flow and transport solvers (in seconds per timestep, SPT), along with their respective parallel efficiencies. Figure~\ref{fig:StrongScal} presents the strong scaling trends for both domains. In an ideal case, the execution time decreases linearly with the number of cores, resulting in a parallel efficiency of one. Departures from this ideal reflect the impact of communication overheads, load imbalance, and algorithmic limitations. The dashed line indicates the limit beyond which the architecture cannot scale to the full image while remaining within the 1,000-node allocation limit.

The mesh generation stage maintains consistently high efficiency across the full range of core counts. The efficiency remains stable, even beyond 10,000 cores. This behaviour highlights the limited communication requirements during mesh construction, which remains mostly local once the domain has been decomposed. Although parallel mesh generation introduces additional data structures such as \texttt{procAddressing} and \texttt{procBoundary}, these are initialised once and do not grow with the number of cores, allowing performance to remain consistent at scale.

At low to moderate core counts, the solvers exhibit superlinear increase in parallel efficiency. This behaviour is primarily attributed to cache effects and improved memory utilisation. As the problem is distributed over more processors, the working set on each core becomes smaller and can increasingly fit within the cache hierarchy, reducing memory-access latency and improving arithmetic throughput. However, as the number of cores continues to grow, the number of cells per processor eventually becomes too small to offset the communication and synchronisation overheads introduced by parallelisation. Beyond this point, the cost of interprocess communication, especially during collective operations such as pressure correction, multigrid coarsening, and Krylov solver reductions, begins to dominate \cite{doi:10.1177/1094342020925535}. Consequently, efficiency starts to decline as computation per processor becomes insufficient to mask communication latency. This transition marks the practical limit of strong scaling for a given problem size. However, in our case, this limit occurs well beyond the maximum allocation available on ARCHER2.

The results demonstrate that our workflow scales efficiently up to the allocation limit. We therefore select the largest configurations within this limit, i.e. 512 cores for Bentheimer and 2,048 cores for Estaillades, to serve as the starting points for the weak scaling analysis, which extends to the full image sizes.

\subsection{Weak Scaling Analysis}
\label{sec:weakScal}

\begin{table}[!t]
\centering
\caption{Weak scaling configurations for the Bentheimer case. Each row corresponds to a configuration with a fixed number of cells per core (371,338). Computational time is reported in seconds per time-step (STP). Parallel efficiency is calculated with respect to the 1-node configuration.}
\label{tab:scal_bentheimer}
\begin{tabular}{cccccccccccccccc}
\toprule
\textbf{NX} & \textbf{NY} & \textbf{NZ} & \textbf{Cores} & \textbf{Node} & \multicolumn{2}{c}{\textbf{Mesh Gen.}}  & \multicolumn{2}{c}{\textbf{Flow (GAMG)}} & \multicolumn{2}{c}{\textbf{Flow (PBiCGStab)}} & \multicolumn{2}{c}{\textbf{Transport}}  \\
& & & & & \textbf{Time (s)} & \textbf{Eff.} & \textbf{SPT} & \textbf{Eff.} & \textbf{SPT} & \textbf{Eff.} & \textbf{SPT} & \textbf{Eff.} \\
\midrule
975 & 975 & 50 & 128 & 1 & 12.5 & 1.00 & 1.41 & 1.00 & 5.41 & 1.00 & 2.11 & 1.00 \\
1,950 & 975 & 50 & 256 & 2 & 12.4 & 1.01 & 1.43 & 0.99 & 5.45 & 0.99 & 2.15 & 0.98 \\
1,950 & 1,950 & 50 & 512 & 4 & 12.6 & 0.99 & 1.49 & 0.95 & 5.48 & 0.99 & 2.17 & 0.97 \\
1,950 & 1,950 & 100 & 1,024 & 8 & 12.7 & 0.98 & 1.46 & 0.97 & 5.47 & 0.99 & 2.13 & 0.99 \\
1,950 & 1,950 & 200 & 2,048 & 16 & 12.7 & 0.98 & 1.52 & 0.93 & 5.65 & 0.96 & 2.20 & 0.96 \\
1,950 & 1,950 & 400 & 4,096 & 32 & 12.5 & 1.00 & 1.55 & 0.91 & 5.66 & 0.96 & 2.18 & 0.97 \\
1,950 & 1,950 & 800 & 8,192 & 64 & 12.6 & 0.99 & 1.57 & 0.90 & 5.68 & 0.95 & 2.20 & 0.96 \\
1,950 & 1,950 & 1,600 &  16,384 & 128 & 12.5 & 1.00 & 1.61 & 0.88 & 5.71 & 0.95 & 2.22 & 0.95 \\
1,950 & 1,950 & 3,200 & 32,768 & 256 & 12.4 & 1.01 & 1.70 & 0.83 & 5.80 & 0.93 & 2.23& 0.95 \\
1,950 & 1,950 & 6,400 & 65,536 & 512 & 12.6 & 0.99 & 2.09 & 0.67 & 6.10 & 0.89 & 2.35 & 0.90 \\
1,950 & 1,950 & 10,800 & 110,592 & 864 & 12.5 & 1.00 & 2.90 & 0.49 & 6.30 & 0.86 & 2.45 & 0.86 \\
\end{tabular}
\end{table}

\begin{table}[!t]
\centering
\caption{Weak scaling configurations for the Estaillades case. Each row corresponds to a configuration with a fixed number of cells per core (95,855). Computational time is reported in seconds per time-step (STP). Parallel efficiency is calculated with respect to the 1-node configuration.}
\label{tab:scal_estaillades}
\begin{tabular}{cccccccccccccc}
\toprule
\textbf{NX} & \textbf{NY} & \textbf{NZ} & \textbf{Cores} & \textbf{Nodes} & \multicolumn{2}{c}{\textbf{Mesh Gen.}}  & \multicolumn{2}{c}{\textbf{Flow (GAMG)}} & \multicolumn{2}{c}{\textbf{Flow (PBiCGStab)}} & \multicolumn{2}{c}{\textbf{Transport}}  \\
& & & & & \textbf{Time (s)} & \textbf{Eff.} & \textbf{SPT} & \textbf{Eff.} & \textbf{SPT} & \textbf{Eff.} & \textbf{SPT} & \textbf{Eff.} \\
\midrule
572 & 286 & 75 & 128 & 1 & 3.84 & 1.00 & 0.33 & 1.00 & 1.17 & 1.00 & 0.47 & 1.00 \\
572 & 572 & 75 & 256 & 2 & 3.95 & 0.97 & 0.34 & 0.97 & 1.18 & 0.99 & 0.47 & 1.00 \\
1,144 & 572 & 75 & 512 & 4 & 3.85 & 1.00 & 0.35 & 0.94 & 1.19 & 0.98 & 0.47 & 1.00 \\
1,144 & 1,144 & 75 & 1,024 & 8 & 3.86 & 0.99 & 0.36 & 0.92 & 1.20 & 0.97 & 0.48 & 0.98 \\
1,144 & 1,144 & 150 & 2,048 & 16 & 3.95 & 0.97 & 0.37 & 0.89 & 1.21 & 0.97 & 0.48 & 0.98 \\
1,144 & 1,144 & 300 & 4,096 & 32 & 3.86 & 0.99 & 0.38 & 0.87 & 1.22 & 0.96 & 0.48 & 0.98 \\
1,144 & 1,144 & 600 & 8,192 & 64 & 3.88 & 0.99 & 0.41 & 0.80 & 1.23 & 0.95 & 0.49 & 0.96 \\
1,144 & 1,144 & 900 & 12,288 & 96 & 3.88 & 0.99 & 0.44 & 0.75 & 1.25 & 0.94 & 0.49 & 0.96 \\
1,144 & 1,144 & 1,500 & 20,480 &160 & 3.85 & 1.00 & 0.50 & 0.66 & 1.25 & 0.94 & 0.50 & 0.94 \\ 
1,144 & 1,144 & 3,000 & 40,960 &320 & 3.81 & 1.01 & 0.56 & 0.59 & 1.28 & 0.91 & 0.51 & 0.92 \\ 
1,144 & 1,144 & 6,000 & 81,920 &640 & 3.87 & 0.99 & 0.67 & 0.49 & 1.33 & 0.88 & 0.52 & 0.90 \\ 
\end{tabular}
\end{table}

The objective of the weak scaling analysis is to evaluate how well the system handles increasing problem sizes while maintaining a constant number of cells per core, as both the core count and problem size increase in tandem. We start from configurations selected in the strong scaling analysis and then adjust both domain size and core count to explore how the system performs under different scaling scenarios, while ensuring we stay within our 1,000-node limit on ARCHER2.

For each case, we chose the largest configuration from the strong scalability study that remains within our allocation. For Bentheimer, this corresponds to 371,338 cells per core, while for Estaillades, it corresponds to 95,855 cells per core. These configurations result in 110,592 cores, i.e. 864 nodes, for the full Bentheimer image (1,950x1,950x10,800) and 81,920 cores, i.e. 640 nodes for the full Estaillades image (1,144×1,144×6,000), and fit within the 1,000-node limit on ARCHER2.

Once the configurations are selected, we proceed by varying both the number of cores and the domain size, increasing and decreasing these parameters while maintaining the constant number of cells per core. This approach enables parallel efficiency to be assessed across different configurations for both cases.

\begin{figure}[!t]
\centering
\includegraphics[width=\linewidth]{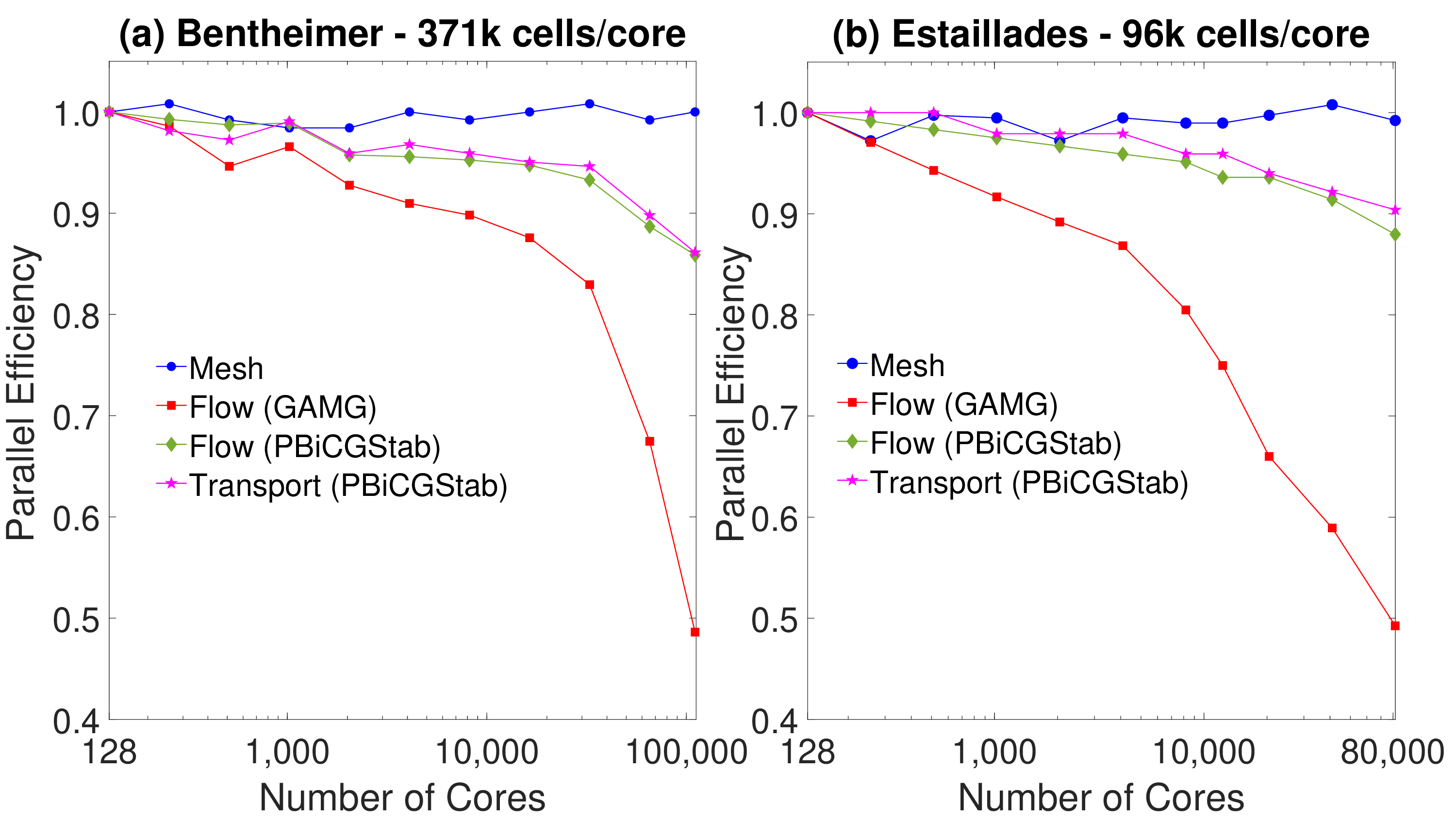}
\caption{Weak scaling results for the Bentheimer (371,338 cells per core) and Estaillades (95,855 cells per core) cases with increasing numbers of cores. Parallel efficiency is calculated with respect to the 1-node configuration.}
\label{fig:weakScalability}
\end{figure}

Tables~ \ref{tab:scal_bentheimer} and \ref{tab:scal_estaillades} present the execution times and corresponding parallel efficiencies for each configuration and each stage of the simulation. Figure~\ref{fig:weakScalability} illustrates the results.

The weak scaling results demonstrate that both the Bentheimer and Estaillades cases maintain high efficiency across most configurations, confirming that the workflow scales well to tens of thousands of cores. Mesh generation shows near-ideal scalability, with efficiency consistently above 0.95 even at the largest core counts. This behaviour reflects the fully distributed nature of the meshing algorithm, where each processor constructs its local portion of the domain independently with minimal interprocess communication.

For the solver stages, performance gradually departs from ideal scaling as the number of cores increases. The PBiCGStab solver maintains relatively stable efficiency, remaining above 0.9 up to the largest tested configurations (65,536 cores for Bentheimer and 81,920 cores for Estaillades). In contrast, the GAMG solver exhibits a stronger decline in scalability, with efficiency dropping from nearly 1.0 at small scales to around 0.5 at the largest. This reduction is caused by the increasing cost of communication and synchronisation across a growing number of processors. Multigrid solvers such as GAMG involve repeated global operations, particularly during coarse-grid correction and interlevel data exchanges, which become increasingly expensive on large distributed-memory systems \cite{doi:10.1177/1094342020925535}.

Despite the reduced scalability at extreme core counts, GAMG remains more computationally efficient than PBiCGStab in absolute terms. Across all configurations, the time per time-step (SPT) for GAMG is at least twice lower than for PBiCGStab, meaning that even with lower parallel efficiency, the total runtime is still shorter. In other words, GAMG achieves faster convergence per step, while PBiCGStab trades some speed for more consistent scaling.

In summary, the weak scaling analysis confirms that the workflow remains efficient at large scales, with near-ideal performance for mesh generation and robust scalability for both solvers up to a hundred of thousands cores. While the PBiCGStab solver achieves more stable efficiency, GAMG remains significantly faster in terms of computational time per step. The solvers scales to the full image with a parallel efficiency of approximatively 50\% for GAMG and 90\% for PBiCGStab.

\section{Ultra-Large Simulations}

Building on the configurations identified in the weak scaling study, we now perform full-core simulations of flow and transport in two representative rock samples: Bentheimer (1,950×1,950×10,800 voxels and Estaillades (1,144×1,144×6,000 voxels). Both domains are discretised at the voxel scale, yielding approximately 41 and 8 billion cells, respectively. Based on the optimal load balancing established earlier, we assign 864 nodes (i.e., 110,592 cores at 371,338 cells per core) for Bentheimer and 640 nodes (i.e. 81,920 cores at 95,855 cells per core) for Estaillades.

We simulate steady-state flow until convergence and perform transient solute transport until 0.5 pore volumes have been injected. The total wall-clock time for the Bentheimer simulation is approximately 50,742 seconds (12 s for mesh generation, 2,230 s for flow, 23,400 s for transport, and 25,100 s for I/O). The Estaillades simulation n takes approximately 28,134  seconds (4 s for mesh generation, 1,230 s for flow, 15,900 s for transport, and 11,000 s for I/O). Table~\ref{tab:large_sims} summarizes the configurations and timing details. While I/O constitutes a significant fraction of the total time, previous analyses suggest variability in these values, so they should be interpreted with caution.

\begin{table}[!t]
\centering
\caption{Simulation configurations and timings for full-core transport in Bentheimer and  Estaillades samples.}
\label{tab:large_sims}
\begin{tabular}{cccccccccccc}
\toprule
& \textbf{NX} & \textbf{NY} & \textbf{NZ} & \textbf{Cores} & \textbf{Nodes}  & \textbf{Mesh Gen. (s)} & \textbf{Flow (s)} & \textbf{Transport (s)} & \textbf{I/O (s)} \\
\midrule
Bentheimer & 1,950 & 1,950 & 10,800 & 110,592 & 864 & 12.5 & 2,230 & 23,400 & 25,100 \\
Estaillades & 1,144 & 1,144 & 6,000 & 81,920 & 640 & 3.87 & 1,230 & 15,900 & 11,000 \\
\end{tabular}
\end{table}

\begin{figure}[!t]
    \centering
    \includegraphics[width=1.0\linewidth]{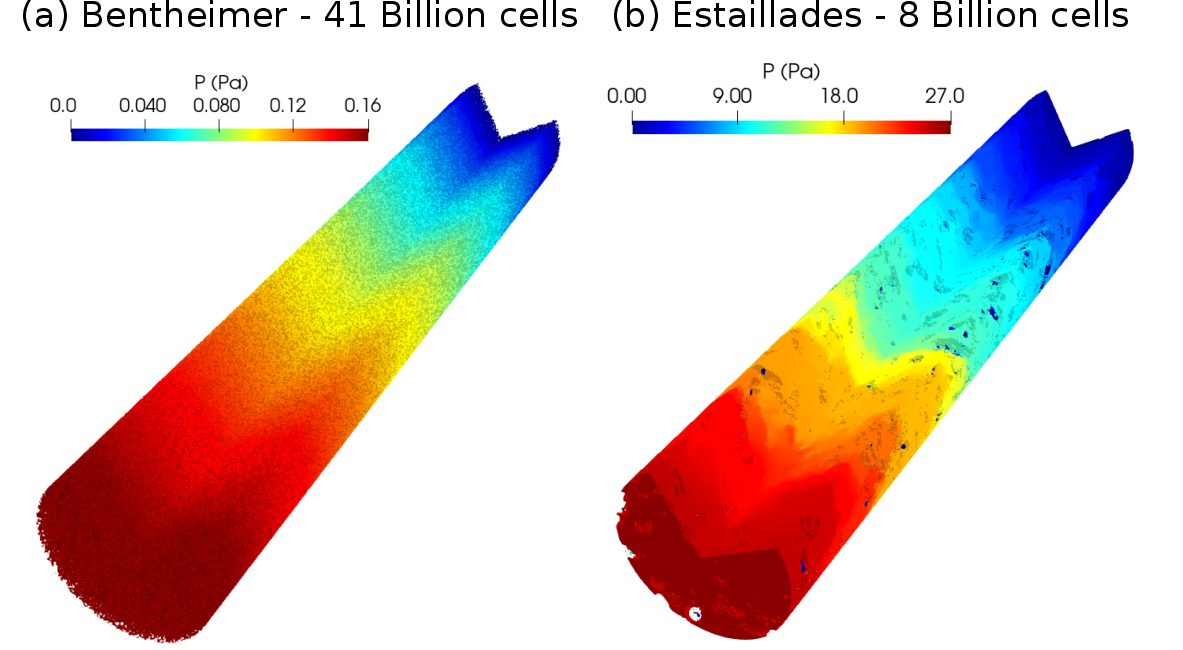}
    \caption{Pressure distribution inside the cylindrical core samples, shown by cropping the top-right quadrant to reveal internal fields. (a) Bentheimer and (b) Estaillades.}
    \vspace{0.5cm}
    \label{fig:Pressure}
\end{figure}

\begin{figure}[!t]
    \centering
    \includegraphics[width=1.0\linewidth]{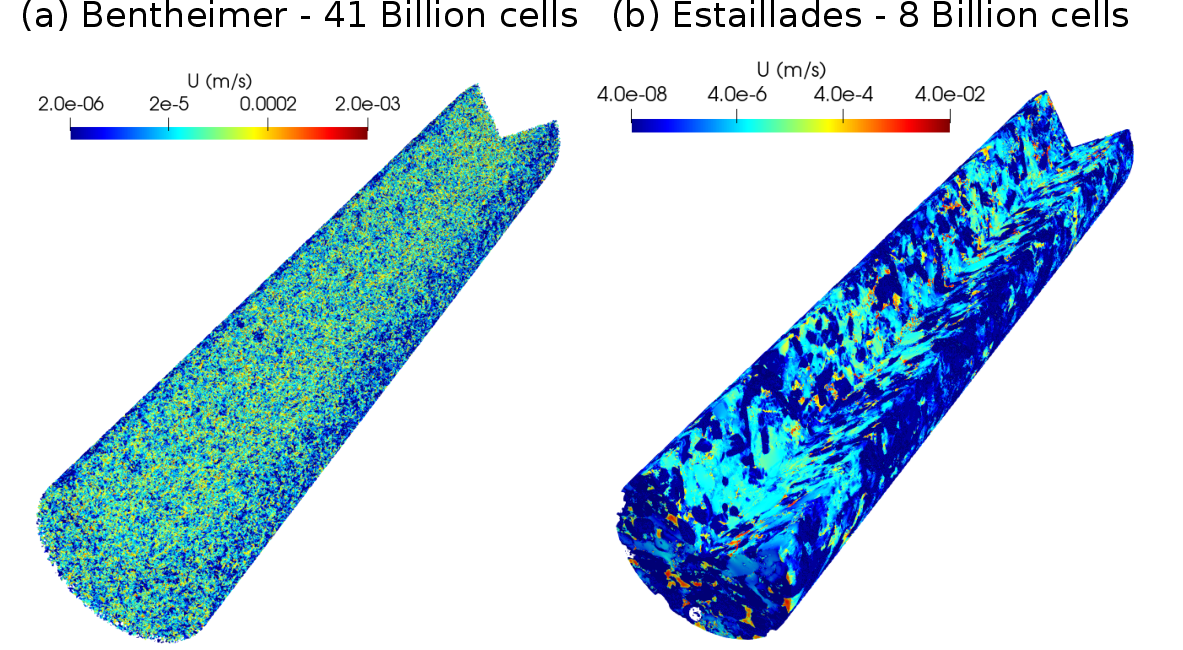}
    \caption{Velocity distribution inside the cylindrical core samples, shown by cropping the top-right quadrant to reveal internal fields. (a) Bentheimer and (b) Estaillades.}
    \label{fig:Velocity}
\end{figure}

Figure~\ref{fig:Pressure} presents the internal pressure fields, visualised by cropping the top-right quadrant of each cylindrical domain. Despite differences in absolute permeability, both samples exhibit a smooth global pressure gradient under steady-state flow. However, more pronounced localised variations are visible in the Estaillades sample due to its greater heterogeneity. The velocity fields in Figure~\ref{fig:Velocity} underscore the contrasting pore-scale structures. Bentheimer exhibits a relatively uniform velocity field, consistent with its homogeneous sandstone microstructure. In contrast, Estaillades shows a much broader range of velocities, influenced by high-permeability macropores interspersed with nearly stagnant microporous regions.

\begin{figure}[!t]
    \centering
    \includegraphics[width=1.0\linewidth]{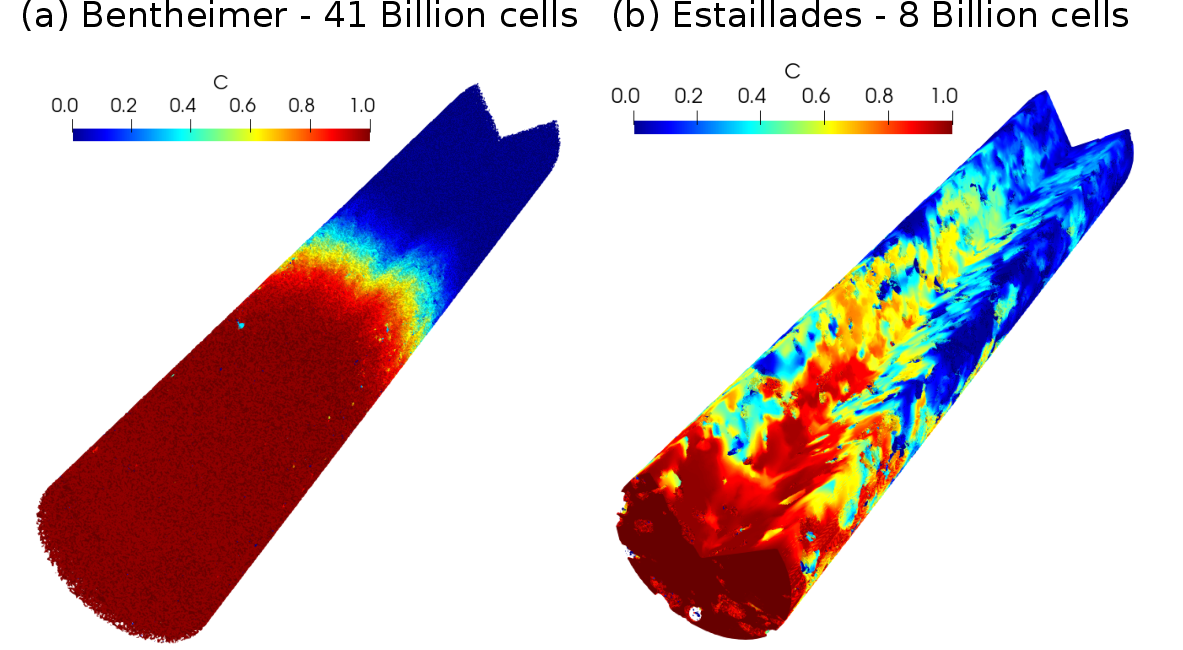}
    \caption{Solute concentration distribution after injection of 0.5 pore volumes, with internal fields revealed by cropping the top-right quadrant. (a) Bentheimer and (b) Estaillades.}
    \label{fig:concentration}
\end{figure}

\begin{figure}[!t]
    \centering
    \includegraphics[width=1.0\linewidth]{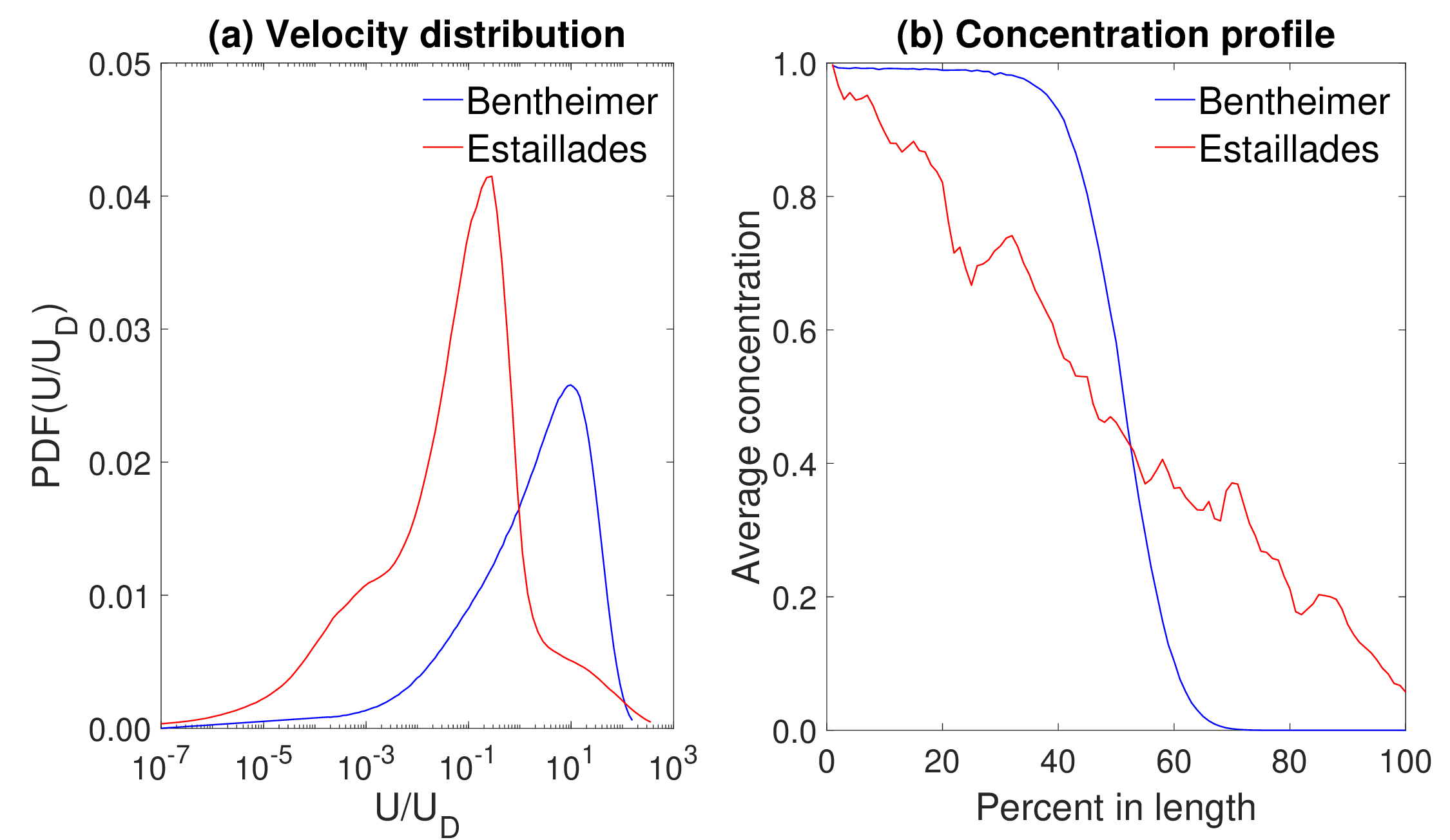}
    \caption{(a) Velocity distribution within the pore space, normalised by the Darcy velocity. (b) Average solute concentration along the core axis after 0.5 pore volumes injected.}
    \label{fig:profile}
\end{figure}

These flow characteristics significantly influence solute transport. Figure~\ref{fig:concentration} shows the concentration fields after injection of 0.5 pore volumes. In Bentheimer, the solute front remains relatively sharp, indicating limited hydrodynamic dispersion and consistent with a Péclet number near unity. In contrast, the Estaillades sample exhibits widespread plume spreading, driven by the heterogeneous velocity field. These differences are further quantified in Figure~\ref{fig:profile}. Panel (a) shows the velocity magnitude distributions, scaled by the Darcy velocity, confirming the broader spread in Estaillades. Panel (b) plots the average concentration along the core length, highlighting the sharp front in Bentheimer and the more gradual transition in Estaillades, indicative of enhanced dispersion.

\begin{figure}[!t]
    \centering
    \includegraphics[width=0.9\textwidth]{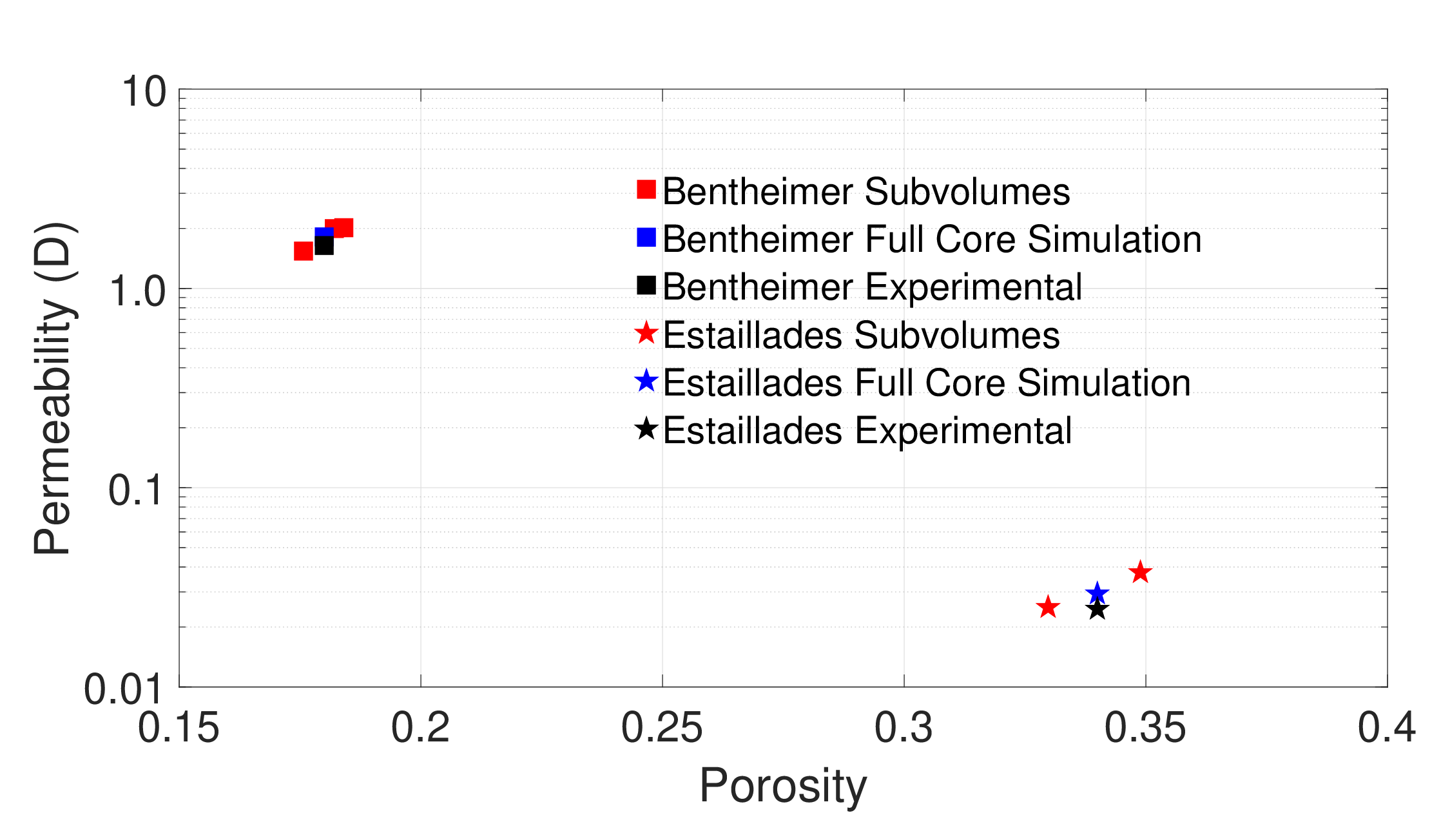}
    \caption{Permeability versus porosity for Bentheimer and Estaillades samples. Blue markers represent full-core simulation results, red markers indicate subvolumes, and black markers show experimental data.}
    \label{fig:poro_perm}
\end{figure}

Figure~\ref{fig:poro_perm} compares the simulated permeabilities with experimental values—taken from \cite{Jackson2020} for Bentheimer and \cite{2022-Menke} for Estaillades. The results show good agreement, with an error of 10.4\% for Bentheimer and 19.5\% for Estaillades. Porosity, derived directly from image data, matches the experimental values.

To investigate the effects of domain size, we also computed permeability on smaller subvolumes. Bentheimer was partitioned into three equal sections along the z-axis (1,950×1,950×3,600 voxels each), while Estaillades was halved (1,144×1,144×3,000 voxels). The porosity and permeability for these subdomains are also shown in Figure~\ref{fig:poro_perm}. Although subvolume simulations remain in reasonable agreement with experiments, they exhibit increased error—up to 2.82\% in porosity and 22.6\% in permeability for Bentheimer, despite the homogeneous nature of the sample. For Estaillades, the error goes up to 3.0\% for porosity and 52.5\% for permeability. These discrepancies, especially in the more heterogeneous sample, underscore the importance of ultra-large domains for accurate characterization.

These results demonstrate the ability of our solver to efficiently perform high-fidelity simulations of flow and transport in ultra-large, highly heterogeneous porous media, using parallel configurations optimised for weak scaling. The good agreement with experimental data confirms the robustness of the approach and its applicability to realistic, high-resolution modelling of porous systems.

\section{Conclusion}

We have presented a scalable and efficient workflow for simulating single-phase flow and solute transport in segmented 3D rock images, based on an immersed boundary approach that avoids the need for generating conformal, body-fitted meshes. While our implementation is demonstrated within the Darcy-Brinkman-Stokes framework, the core methodology—relying on voxel-based Cartesian grids and avoiding explicit reconstruction of fluid–solid interfaces—can be extended to any immersed boundary method that supports this representation. This flexibility enables a robust and generalizable framework for high-resolution pore-scale modelling across a range of physical formulations.

The structured mesh architecture allows for straightforward domain decomposition, where each subdomain is processed independently with minimal communication overhead. This design significantly simplifies the parallel implementation and proves highly effective for large-scale simulations. Compared to standard unstructured meshing workflows, our approach significantly reduces preprocessing time and memory consumption, making it more suitable for extreme-scale applications. Our performance evaluation confirms that the full simulation pipeline, including mesh generation, flow, and transport, scales efficiently to tens of thousands of cores. In particular, mesh generation demonstrates ideal scalability, if we ignore the I/O times which varied significantly on the shared system. While flow and transport solvers maintain consistent performance at large core counts, up to approximatively 100,000 cores, keeping a parallel efficiency of at least 50\%.

We applied the method to simulate flow and transport in two large, realistic rock samples—Bentheimer sandstone and Estaillades limestone—using up to 41 billion cells. The simulations captured sharp solute fronts in the homogeneous Bentheimer sample and extensive dispersion in the more heterogeneous Estaillades sample, reflecting their underlying pore structures. Comparisons with experimental measurements show that our full-core simulations achieve good agreement: permeability errors are within 10.4\% for Bentheimer and 19.5\% for Estaillades. Subvolume analyses, however, reveal significantly higher deviations—up to 22.6\% in permeability for Bentheimer and 52.5\% for Estaillades—emphasizing the importance of modeling at sufficiently large scales to capture representative behavior in heterogeneous media.

Overall, the workflow enables direct, high-fidelity simulations of flow and transport in complex porous media without preprocessing or geometric simplification. It offers a practical and scalable tool for digital rock physics, uncertainty quantification, and the interpretation of experimental core-scale data.

Future work will focus on extending this framework to support multiphase flow and reactive transport. These additions will introduce further complexity, including non-linear coupling between phases, phase interface dynamics, and potentially stiff chemical reactions. Addressing these challenges will require continued optimization of solver components, adaptive strategies to manage localised complexity, and support for emerging computing architectures, including GPU acceleration. In the long term, these developments aim to broaden the scope of image-based modelling to tackle increasingly realistic and multiscale porous media systems in both scientific and industrial applications.


\section*{Acknowledgement}
This work was funded under the eCSE ARCHER2 program, the EPSRC ECO-AI grant (references: EP/Y006143/1, EP/Y005732/1) and the NWS-NERC GeoSAFE grant (reference NE/Y002504/1). Opinions, findings, conclusions or recommendations expressed are those of the authors and do not necessarily reflect the views of EPSRC, NERC or NWS.

\bibliographystyle{spphys} 

\bibliography{cas-refs}

\end{document}